# BEAM STUDIES FOR THE PROTON IMPROVEMENT PLAN (PIP) - REDUCING BEAM LOSS AT THE FERMILAB BOOSTER *

K. Seiya#, C. M. Bhat, D. E. Johnson, V. V. Kapin, W. A. Pellico, C. Y. Tan, R. J. Tesarek, FNAL, Batavia, IL 60510, USA

*Abstract*

The Fermilab Booster is being upgraded under the Proton Improvement Plan (PIP) to be capable of providing a proton flux of 2.25E17 protons per hour. [1] The intensity per cycle will remain at the present operational 4.3E12 protons per pulse, however the Booster beam cycle rate is going to be increased from 7.5 Hz to 15 Hz. One of the biggest challenges is to maintain the present beam loss power while the doubling the beam flux. Under PIP, there has been a large effort in beam studies and simulations to better understand the mechanisms of the beam loss. The goal is to reduce it by half by correcting and controlling the beam dynamics and by improving operational systems through hardware upgrades. This paper is going to present the recent beam study results and status of the Booster operations.

## 700KW OPERATION

Fermilab is going to provide 700kW proton beam to the NOvA experiment. Booster is a 15 Hz resonant circuit synchrotron and accelerates proton beams from 400 MeV to 8 GeV. The required intensity in the Booster for NOvA is 4.3E12 ppp, the same as it was for 400 kW operation. [2] However, the cycle rate will be increased from ~7.5 Hz to 15 Hz to accommodate both NOvA and other users. The RF system and utilities are being upgraded to 15 Hz operations and are nearing completion. The plan is to start 15 Hz operations in FY15.

The beam loss limit has been set to 525W to allow workers to maintain all elements in the Booster tunnel without excessive radiation exposure. Figure 1 shows the historical beam loss in the Booster versus protons per hour. The total loss depends on the beam intensity.

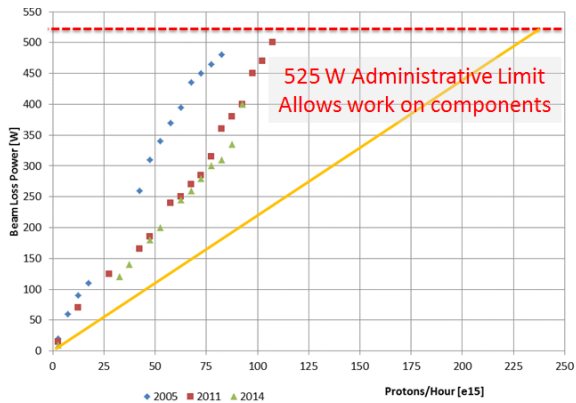

Figure 1: Beam power loss for 3 year operations (blue: 2005, red: 2011 and green: 2014).

___________________________________________
*Work supported by Fermilab Research Alliance, LLC under Contract No. DE-AC02-07CH11359 with the United States Department of Energy.
#kiyomi@fnal.gov

The present operational beam intensity at injection is about 5E12 ppp and extraction is 4.5E12 ppp. The total energy loss is 0.075 kJ in one Booster cycle and hence 1150 W when the cycle rate is 15Hz. The loss has to be reduced to half by 2016. Figure 2 shows the intensity and loss during normal operations. Beam studies and upgrades that will be done to reduce the beam losses will be discussed in this paper.

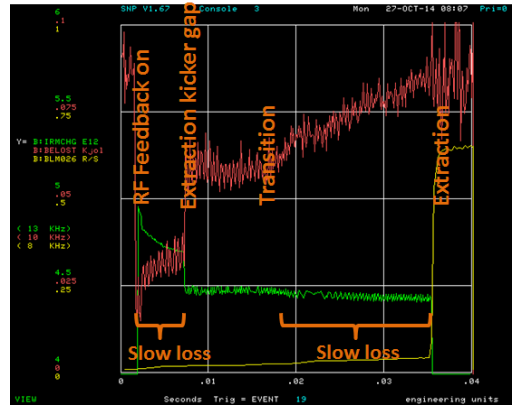

Figure 2: Beam intensity (green), energy loss (red) and extraction loss (yellow) signals.

## COLLIMATORS

Two-stage collimator system which had horizontal and vertical primary collimators and three secondary collimators was installed in Booster in 2004. The primary collimators haven't been used and the secondary collimators have been used as single-stage collimators because the system never worked or implemented as designed which created much larger scatter than we expected from the simulation results.

Ramped corrector magnets, installed in the Booster during the 2007/2009 shutdowns, allowed us to control the beam orbit. In 2013, realignment of the main dipole magnets based on aperture scans and simulation increased aperture. RF Cogging changed the radial positon from cycle to cycle and it was replaced with Magnetic Cogging in 2015 which kept beam on central orbit. With the better understanding and control of the orbit, the operating parameters for the collimators will be optimized based on simulations and measurements.

A simulation was developed using MADX to replace an earlier STRUCT simulation and better optimize the material and thickness of the primary collimators. Assuming that $10^4$ particles at the edge of horizontal primary collimators with copper foil of 381μm and 50μm thickness, the halo particles were simulated. Figure 3 shows the distribution of the particle loss at 400MeV along the ring. The collimation energy and beam orbit will be optimized using the simulations.

Two beam loss monitors were developed with two PMT and 4 scintillators on each monitor and installed near the vertical primary collimator and at the second secondary collimator where the normal loss monitors are located. Figure 4 shows that the new loss monitor is able to measure beam loss within one rf bucket.

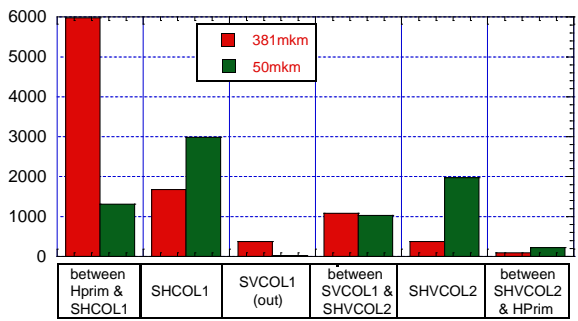

Figure 3: Beam loss pattern around the Booster ring. The horizontal collimator was located at '0'.

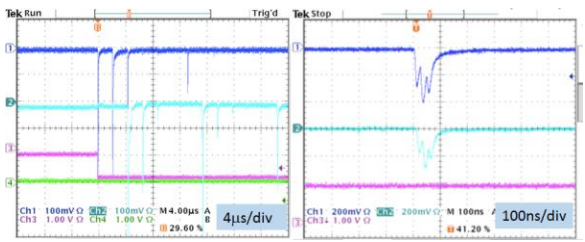

Figure 4: Output signals of PMT1(blue) and PMT3 (cyan) when 3 bunches are kicked out with Notcher. The measurements taken here were when the notch was formed at 5200us into the cycle. At this point, rev = 1.9us and RF = 1/44MHz ~23ns.

## BEAM INJECTION AND CAPTURE

Two hundred MHz bunches are injected from LINAC with multi turn injection and they are adiabatically captured with 37.9 MHz RF voltage. Uncaptured particles are going to be beam loss when the RF feedback is turned on.

Beam energy spread was measured at 400 MeV Booster injection [3] by creating a gap of ≈40 μsec in the coasting beam and measuring the velocity of the beam slipped into the gap. The energy spread of the beam was also estimated by using multi wire profile monitors data at the end of the LINAC and comparing the results with calculated lattice. The total energy spread was 1.50±0.16 MeV.

A new injection scheme called 'Early Beam Injection Scheme' (EIS) [4] has been proposed to reduce the losses in the Booster cycle. In this scheme, the injection time is moved earlier than 'Current Injection Scheme' (CIS) by

≈150 μsec as shown in Figure 5. The injected beam in EIS stays near 400 MeV until the capture is completed. Both schemes were simulated with the measured energy spread and the predicted longitudinal phase space distributions just before acceleration are compared in Figure 6. Simulation results show that 2% of beam loss and 50% of emittance blow up with the CIS while no particles outside of the RF bucket and no emittance dilution with the EIS.

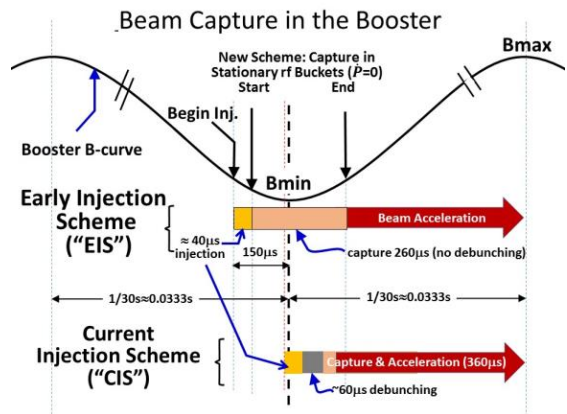

Figure 5: New injection scheme and current scheme.

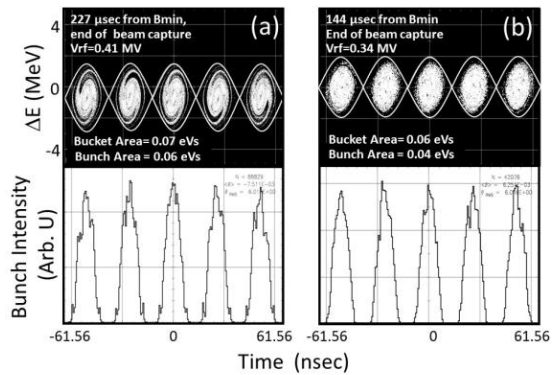

Figure 6: Simulated longitudinal phase space distributions just before the acceleration (a) current scheme and (b) early injection scheme.

## OPTICS CORRECTION WITH LOCO

The beta functions are distorted at injection because of the DC extraction bump magnets. The modulation could cause emittance mismatch and induce unexpected instabilities. The ideal lattice was set using MADX and minimized the distortion. LOCO (Linear Optics from Closed Orbit) calculates the quad and skew quad currents that corrects the measured lattice to make it closer to the ideal lattice. [5]

The corrected lattice was applied on a study cycle and it ran beam with a reasonable efficiency. It is planned to have this lattice to be used in operations in the near future.

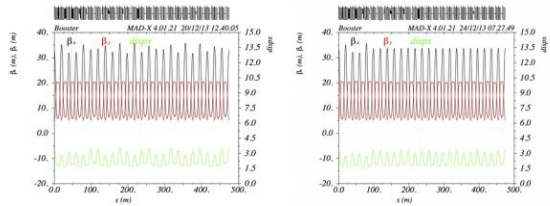

Figure 7: The calculated lattice with the dogleg turned on (left) and the corrected lattice (right).

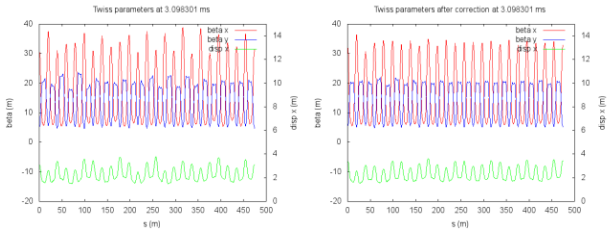

Figure 8: The measured Booster lattice at injection before (red) and after (blue) the correction in the horizontal (left) and vertical (right) planes.

## NOTCH AND COGGING

The beam fills Booster circumference after the injection capture if there is no gap in the LINAC pulse. The extraction kicker gap called 'Notch' must be created at low energy to allow the 8 GeV extraction kicker to fire without creating losses.

### *Magnetic cogging*

New beam extraction synchronization system called "Magnetic Cogging" was developed at the Fermilab Booster. The main dipole field error causes variation of the revolution frequency pattern during the cycle and changes the final position of the extraction bucket from cycle to cycle. The Magnetic Cogging controls the position of the extraction kicker gap by changing the dipole corrector fields. The feedback system was built with a new programmable VXI board and the gain was optimized using the simulation results. The gap creation was used to be at 700 MeV with the RF Cogging and it was moved to 400 MeV with Magnetic Cogging. The system was successfully implemented to the operational in March 2015. The Magnetic cogging reduced beam energy loss at the Notch creation by 40 % and the total cycle loss by 15%.

### *Laser Notch*

A laser system is being built to create the notch within a LINAC beam pulse, immediately after the RFQ at 750 keV, where activation issues are negligible. (Figure 9) [7] The laser system is a MOPA design with three stages of fiber and two stages of solid state amplification. This will create a burst of spatially and temporally uniform 200 MHz pulses, each with 2 mJ of energy, to match ion bunch structure out of the RFQ to create a set of notches in the linac pulse at the Booster revolution period. A demonstration experiment to verify neutralization efficiency using the installed optical cavity was performed earlier this year. The single bunch neutralization at different laser pulse energies is shown in Figure 10.

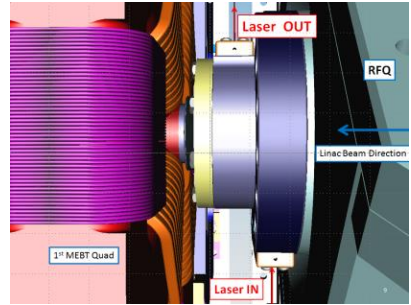

Figure 9: Laser notching cavity installed on the downstream RFQ flange.

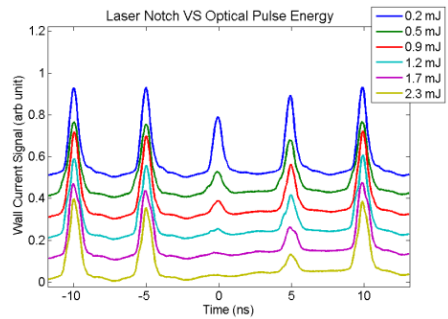

Figure 10: Wall current monitor signal in the LINAC with 6 different optical pulse energies.

## SUMMARY

The Booster cycle rate of 15 Hz and averaging 4.3E12 protons per pulse will be completed by 2016. The beam loss has to be reduced by half compared to the present situation. The ongoing PIP beam studies along with hardware and software upgrades are critical. A successful completion of the PIP effort is a laboratory priority and essential for reaching the HEP proton delivery goals.

## REFERENCES

[1] W. Pellico et al., "FNAL - The Proton Improvement Plan (PIP)", THPME075, IPAC'14, Dresden, Germany (2014).
[2] I. Kourbanis, "Progress Towards Doubling the Beam Power at Fermilab's Accelerator Complex", TUOAA01, IPAC'14, Dresden, Germany (2014).
[3] C. M. Bhat, et al., (this conference, THPF113).
[4] C. M. Bhat, (this conference, THPF112).
[5] C. Y. Tan, et al., (this conference, MOPMA020).
[6] K Seiya et al., "Development of Cogging at the Fermilab Booster ", WEPR069, IPAC'14, Dresden, Germany (2014).
[7] D. E. Johnson, et al., (this conference, WEPTY028).